\documentclass[12pt, draftclsnofoot, journal, letterpaper, onecolumn, oneside]{IEEEtran}
\usepackage{cite}
\usepackage{graphicx}
\usepackage{subfigure}
\usepackage{amsmath}
\usepackage{amssymb}
\usepackage{mathtools}
\usepackage{bm}
\usepackage{epsfig}
\usepackage{times}
\usepackage{color}
\usepackage{url}
\usepackage{array}
\usepackage{multirow}
\usepackage{rotating}
\usepackage{extarrows}
\usepackage{algorithm}
\usepackage{algorithmic}
\usepackage{mathabx} 
\usepackage{wasysym}

\newtheorem{pavikc}{\textbf{Corollary}}
\newtheorem{pavikl}{\textbf{Lemma}}

\newtheorem{pavikp}{\textbf{Proposition}}

\newcommand{\argmax}{\operatornamewithlimits{argmax}}

\newcommand{\pder}[2][]{\frac{\partial#1}{\partial#2}}
\newcommand{\der}[2][]{\frac{d#1}{d#2}}
\usepackage{tikz}
\usetikzlibrary{arrows,shapes}

\def\hh{\widehat{\mathbf{h}}_e}
\def\Dh{\widehat{\mathbf{D}}}

\title{Secure Transmission in Amplify-and-Forward Diamond Networks with a Single Eavesdropper}
\author{\IEEEauthorblockN{Siddhartha Sarma\IEEEauthorrefmark{1}, Samar Agnihotri\IEEEauthorrefmark{2} and Joy Kuri\IEEEauthorrefmark{1}
	}
	\thanks{S. Sarma and J. Kuri are with the Department of Electronic Systems Engineering, Indian Institute of Science, Bangalore, Karnataka - 560012, India (e-mail: \{siddharth, kuri\}@dese.iisc.ernet.in).}
	\thanks{S. Agnihotri is with the School of Computing and Electrical Engineering, Indian Institute of Technology Mandi, Mandi, Himachal Pradesh - 175001, India (e-mail: samar@iitmandi.ac.in).}
}

\begin{document}

\maketitle

\begin{abstract}
Unicast communication over a network of $M$-parallel relays in the presence of an eavesdropper is considered. The relay nodes, operating under individual power constraints, amplify and forward the signals received at their inputs. In this scenario, the problem of the maximum secrecy rate achievable with AF relaying is addressed. Previous work on this problem provides iterative algorithms based on semidefinite relaxation. However, those algorithms result in suboptimal performance without any performance and convergence guarantees. We address this problem for three specific network models, with \emph{real-valued channel gains}. We propose a novel transformation that leads to convex optimization problems. Our analysis leads to \textit{(i)} a polynomial-time algorithm to compute the optimal secure AF rate for two of the models and \textit{(ii)} a closed-form expression for the optimal secure rate for the other.
\end{abstract}

\section{Introduction}
\label{sec:intro}
Wireless communication, by its inherent broadcast nature, is vulnerable to eavesdropping by illegitimate receivers within communication range of the source. Wyner in \cite{wyner}, for the first time, information-theoretically addressed the problem of secure communication in the presence of an eavesdropper and showed that secure communication is possible if the eavesdropper channel is a degraded version of the destination channel. The rate at which information can be transferred from the source to the intended destination while ensuring complete equivocation at the eavesdropper is termed as \textit{secrecy rate} and its maximum over all input probability distributions is defined as the \textit{secrecy capacity} of channel. Later, \cite{078cheongHellman} extended Wyner's result to Gaussian channels. These results are further extended to various models such as multi-antenna systems \cite{105paradaBlahut, 110khistiWornell}, multiuser scenarios \cite{108liuMaric, 108khisti_thesis}, fading channels \cite{108liangPoorShamai, 108gopalaLaiGamal}.

An interesting direction of work on secure communication in the presence of eavesdropper(s) is one in which the source communicates with the destination via relay nodes \cite{108laiGamal, 109dongHanPetropuluPoor, 110zhangGursoy, 110zhangGursoy2, 113yangLiMaChiang}. Such work has considered various scenarios such as different relaying schemes (\textit{amplify-and-forward} and \textit{decode-and-forward}), constraints on total or individual relay power consumption, one or more eavesdroppers. However, tight characterization of secrecy capacity or even optimal achievable rate is not available except for a few specific scenarios.

We consider unicast communication over a network of $M$ parallel relays in the presence of an eavesdropper. The relay nodes, operating under individual power constraints, amplify and forward signals received at their inputs. The objective is to maximize the rate of secure transmission from the source to the destination by choosing the optimal set of scaling factors for the AF-relays. In general, the problem is non-convex and global optima cannot be guaranteed. However, we establish that convexification of this optimization problem is possible for \textit{degraded eavesdropper channels} scenario where each relay to destination channel gain is larger than the corresponding relay to eavesdropper channel gain. Though such a scenario is limited, we argue that it may still provide insights into the nature of the optimal solution for general problem and help us construct low-complexity capacity achieving achievability schemes.
Later, we discuss two special scenarios: \textit{scaled eavesdropper channels} and \textit{symmetric network} and provide a polynomial-time algorithm and a closed-form solution, respectively, to compute the optimal secure AF rate.

To the best of our knowledge, we are the first to propose a convex optimization based solution to obtain the global optimum for the degraded channel scenario, where channel gains are real numbers\footnote{The more general case with complex channel gains is currently under investigation.}. Previously, the problem addressed in this paper was also considered for general channel scenarios by \cite{110zhangGursoy} for a single eavesdropper and \cite{113yangLiMaChiang} for multiple eavesdroppers. However, the semidefinite relaxation based approaches proposed therein only provide suboptimal performance without any guarantee on the performance (even for simple network models) and the convergence of the proposed iterative algorithms. Unlike such previous work, we propose a novel variable transformation that results in convex optimization for the degraded channel scenario. 

Our main contributions are as follows:
\begin{itemize}
\item For the degraded eavesdropper channel model, we propose a novel transformation for the non-convex problem and present an iterative algorithm with guaranteed convergence to compute the optimal secrecy rate achievable with AF relaying. We also provide a lower-bound to the optimal rate based on the \textit{zero-forcing} approach and an upper bound based on the total relay power constraint.
\item For the network model with scaled eavesdropper channels, motivated by the greedy approaches proposed in  \cite{109jingJafarkhani}-\cite{111agnihotriJaggiChen} to compute the optimal AF rate over diamond networks, we propose a novel efficient scheme to compute the optimal secrecy rate achievable with AF relaying.
\item Finally, for the symmetric network model, we provide a closed form expression for the optimal secrecy rate achievable with AF relaying, along with the corresponding optimal vector of scaling factors.
\end{itemize}

This paper provides some of the key results of our work. Various extensions and detailed discussions can be found in our ArXiv submission \cite{115sarmaAgnihotriKuri}.

\noindent\textit{Organization:} Section~\ref{sec:model} introduces the system model and notation. In Section~\ref{sec:secrate}, we analyze the optimal AF secrecy rate for three specific network models. The performance of proposed approaches is evaluated in Section~\ref{sec:res}. Finally, Section~\ref{sec:concl} concludes the paper.

\section{System Model}
\label{sec:model}
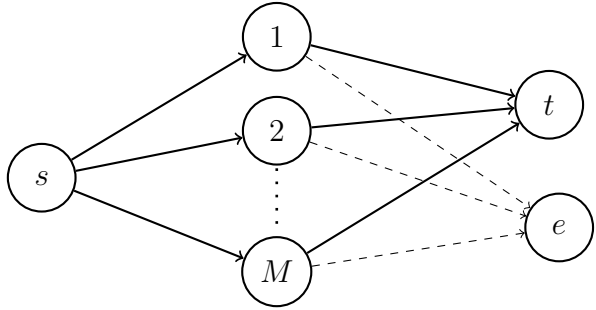
\begin{figure}[!t]
\centering
\begin{tikzpicture}[scale=1.25]
\tikzstyle{every node}=[draw,shape=circle,minimum size=0.9cm,style=thick]
\node (v0) at (180:2.5) {$s$};
\node (v1) at (90:1.5) {$1$};
\node (v2) at (90:0.5) {$2$};
\node (v3) at (270:1) {$M$};
\node (v4) at (15:3) {$t$};
\node (v6) at (350:3.05) {$e$};
\draw[style=thick,->] (v0) -- (v1);
\draw[style=thick,->] (v0) -- (v2);
\draw[style=thick,->] (v0) -- (v3);
\draw[style=thick,->] (v1) -- (v4);
\draw[style=thick,->] (v2) -- (v4);
\draw[style=thick,->] (v3) -- (v4);
\draw[dashed,->](v1) -- (v6);
\draw[dashed,->](v2) -- (v6);
\draw[dashed,->](v3) -- (v6);
\draw[loosely dotted,line width=1](0,0.10) -- (0,-.6);
\end{tikzpicture}
\vspace{-0.1in}
\caption{The diamond $M$-relay AF network with an eavesdropper}
\vspace{-0.2in}
\label{fig:model}
\end{figure}
Consider the Gaussian ``diamond'' or parallel $M$-relay network as depicted in Figure~\ref{fig:model}. The source node $s$ transmits a message to the destination $t$ with the help of $M$ parallel relays. However, the signals transmitted by the relays for the destination are also overheard by the eavesdropper $e$. The channel inputs at time $n \in \mathbb{N}$ at the source $s$ and the $i^{th}$ relay are denoted by $x_s[n]$ and $x_i[n]$, respectively. The channel outputs at time $n \in \mathbb{N}$ at the destination $t$, the eavesdropper $e$, and the $i^{th}$ relay are denoted by $y_t[n], y_e[n]$, and $y_i[n]$. These channel inputs and outputs are related as:
\begin{align}
y_i[n] &= h_{si} x_s[n] + z_i[n] \label{eq:rel} \\
y_k[n] &= \sum_{i=1}^M h_{ik} x_i[n] + z_k[n] \label{eq:destEve},\;k\in\{t,e\}
\end{align}
where $(z_i[n])_{i,n}$, $(z_t[n])_n$, and $(z_e[n])_n$ are independent and identically distributed Gaussian random variables with zero mean and variance $\sigma^2$, independent of channel inputs. The channel gains $(h_{si})_{i=1}^M$, $(h_{it})_{i=1}^M$, and $(h_{ie})_{i=1}^M$ are real numbers, constant over time (\cite{111agnihotriJaggiChen, ursdiggavi}) and known (even for the eavesdropper channels) throughout the network \cite{109dongHanPetropuluPoor,110zhangGursoy,110zhangGursoy2}.

In \textit{amplify-and-forward} relaying, each relay scales its received signal before transmitting. The maximum scaling factor at a relay is determined by its individual power constraint. We consider the following average transmit power constraint:
\begin{equation*}
E[x_i^2[n]] \le P_i,\quad  -\infty <n<\infty,\quad i \in \{s,1,2,\dots,M\}
\end{equation*}
Therefore, the channel input at the $i^{th}$ relay can be written as:
\begin{equation}
\label{eq:beta}
x_i[n+1]=\beta_i y_i[n],
\end{equation}
where the scaling factor $\beta$ is subject to the constraint 
\begin{align}\label{eq:beta_cons}
0 \le \beta_i^2 \le \beta_{i,max}^2 = P_i/(h_{si}^2 P_s + \sigma^2)
\end{align}

Assuming equal delay along each path, for the network in Figure \ref{fig:model}, the copies of the source signal ($x_s[.]$) and noise signals ($z_i[.]$), respectively, arrive at the destination and the eavesdropper along multiple paths of the same delay. Therefore, the signals received at the destination and eavesdropper are free from intersymbol interference (ISI). Thus, we can omit the time indices and use equations \eqref{eq:rel}-\eqref{eq:beta} to write:
\begin{equation}
y_k=\sum\limits_{i=1}^{M} h_{si}\beta_i h_{ik} x_s + \sum\limits_{i=1}^{M}\beta_i h_{ik} z_i + z_k,\;k\in\{t,e\}
\end{equation}
The secrecy rate at the destination for such a network model can be written as \cite{wyner},  
$R_s(P_s)= [I(x_s;y_d)-I(x_s;y_e)]^+$,
where $I(x_s;y)$ represents the mutual information between random variable $x_s$ and $y$ and $[u]^+=\max\{u,0\}$.
The secrecy capacity is attained for Gaussian channels with the Gaussian input $x_s \sim \mathcal{N}(0,P_s)$, where $\mathbf{E}[x_s^2] = P_s$, \cite{078cheongHellman}. Therefore, the optimal secrecy rate achievable with AF relaying can be written as the following optimization problem.
\begin{subequations}
\label{eq:opt1}
\begin{align}
R_s(P_s) &=  \max_{\bm{\beta}} \left[R_d(P_s,\bm{\beta}) - R_e(P_s,\boldsymbol{\beta})\right]\\
         &=  \max_{\boldsymbol{\beta}} \left[\frac{1}{2}\log\frac{1+SNR_d(P_s,\bm{\beta})}{1+SNR_e(P_s,\bm{\beta})}\right],
\end{align}  
\end{subequations}
where $SNR_k(P_s,\bm{\beta}) =  \gamma_s \Gamma_k(\bm{\beta}),\;k \in \{t,e\}$ with
\begin{equation}
\label{eq:snr_vec}
\Gamma_k(\bm{\beta})= \frac{(\mathbf{h_{s,k}}^t\bm{\beta})^2}{1+\bm{\beta}^t\mathbf{D_k}\bm{\beta}}, 
\end{equation}
$\gamma_s = \frac{P_s}{\sigma^2}$, $\mathbf{h_{s,k}}=[h_{s1}h_{1k},h_{s2}h_{2k},\cdots,h_{sM}h_{Mk}]^t$, and 
$\mathbf{D_k}=diag([h_{1k}^2,h_{2k}^2,\cdots,h_{Mk}^2])$ is a $M\times M$ diagonal matrix with diagonal entries written in vector form.

In the following section, we analyze the optimal secure AF rate problem in \eqref{eq:opt1} for three specific network models.

\section{The Optimal AF Secrecy Rate Analysis}
\label{sec:secrate}
We discuss the optimal solution of the maximum AF secrecy rate problem \eqref{eq:opt1} in the $M$-relay diamond networks with single eavesdropper in \textit{degraded eavesdropper channels} scenario. Later we consider two special cases, namely \textit{scaled eavesdropper channels} and \textit{symmetric network}.

\subsection{Degraded Eavesdropper Channels}
\label{subsec:degradedEve}
Consider a general channel model where gains along the source to relays and relays to the destination channels may take arbitrary values. However, the relay to eavesdropper channels are degraded versions of the relay to destination channels, that is, $h_{it}>h_{ie}$ or $\bm{h_e}=\bm{\alpha}^t \bm{h_t}, \alpha_i < 1, i \in \{1, \ldots, M\}$. 
{
\subsubsection{Optimal secrecy rate for individual relay constraints}
From equation \eqref{eq:opt1} we can rewrite an equivalent optimization problem in the following manner:
\begin{subequations}
	\label{eq:optindvcons}
	\begin{align}
		\max_\eta\;\;\max_{\bm{\beta}}\;\;& \frac{1+SNR_d(\bm{\beta})}{1+\eta} \label{objfn:optindv}\\
		\text{subject to:}\;\; & SNR_e(\bm{\beta})\le \eta,\\ 
		& 0 \le \beta_i^2 \le \beta_{i,max}^2, \; i \in \{1,2,\dots M \}
	\end{align}
\end{subequations}	
The optimization problem \eqref{eq:optindvcons} can be solved in two steps. At first, we fix an $\eta$ and solve the inner optimization problem for optimal $\bm{\beta}$. We first present an approach to solve the inner optimization problem, then we address the second step involving the variable $\eta$.

\par We perform the following transformation on the inner optimization \eqref{eq:optindvcons} to convert the non-convex problem into a convex one.
$ v_i={\omega_i}/{(1+\sum_{i=1}^{M}\omega_i^2)^{\frac{1}{2}}}$
\text{or, equivalently } 
$\bm{\omega}=\frac{\mathbf{v}}{\sqrt{1-\mathbf{v^tv}}}$.\\
The objective function becomes $\mathbf{v^th_sh_s^tv}$ and it can be shown that it suffices to maximize the linear objective $\mathbf{h_s^tv}$. The reformulated inner optimization problem is:
\begin{subequations}\label{opt_indv}
	\begin{align}
	\max_\mathbf{v}&\; \mathbf{h_s^tv}\\
	\text{subject to:}&\;\mathbf{v^tC_e}(\eta)\mathbf{v} \le 1,\label{eq:evecons}\\ 
	& \mathbf{v^tD_iv} \le 1, \; i \in \{1,2,\dots M \},\label{eq:consindv}
	\end{align}
\end{subequations}
where $\mathbf{C}_e(\eta)=\frac{\hh\hh^\mathbf{t}}{\eta}\frac{P_s}{\sigma^2}+\mathbf{I}-\Dh_e$, $\hh=[h_{s1}\rho_1,h_{s2}\rho_2,$ $\ldots,h_{sM}\rho_M]^\mathbf{t}$, $\Dh_e=diag([\rho_{1}^2,\rho_{2}^2,\ldots,\rho_{M}^2])$ and $\rho_i=\frac{h_{ie}}{h_{id}},\;\forall i$. 
$\mathbf{D_i}$ characterizes the constraint for each relay node and is given by $$(\mathbf{D_i})_{mn}=
\begin{dcases}
& 1+ \frac{v_i^2}{h_{id}^2\beta_{i,max}^2},\;\text{if }m=n=i\\
& 1 ,\text{ if }m=n\ne i\\
& 0, \text{ otherwise } 
\end{dcases}$$
For the degraded channels $\rho_{i}<1$, so $\mathbf{C}_e(\eta)$ is a positive semidefinite matrix. As the objective function is linear and constraints are convex, so the problem in \eqref{opt_indv} is convex and can be solved optimally using well-known numerical techniques \textit{e.g.} interior-point methods.

Once we have obtained the solution for the inner optimization problem over transformed variable $\bf{v}$, then \eqref{eq:opttotcons} reduces to a one dimensional optimization problem. If we denote the optimal solution for the inner optimization problem by $\bf{v^*}(\eta)$ for a fixed $\eta$, then the following proposition describes the nature of the objection function \eqref{objfn:optindv}. 
\begin{pavikp}
	\label{prpn:unimod}
	The objective function  $f(\eta)=\frac{1+(\mathbf{h_s^tv}^*(\eta))^2}{1+\eta}$ is unimodal in $\eta$ for $\eta \in[0, \infty)$.
\end{pavikp}
\begin{IEEEproof}
	Please refer to Appendix \ref{appnd3}
\end{IEEEproof}
As $f(\eta)$ is unimodal, so we can guarantee the convergence to global optima using line search algorithms (\textit{e.g.} golden section \cite{Chong2013}).
Instead of searching over whole positive real line ($\mathbb{R}^+$), we can limit the range of $\eta$ by further analysis. 

\textit{Range of $\eta$:} As SNR is always positive, so $\eta=0$ serves as a lower bound. This $\eta$ value corresponds to the zero-forcing solution (\ref{subsubsec:zfc}) which serves as a lower bound for optimization problem \eqref{eq:optindvcons}. For upper bound on $\eta$, we solve the following optimization problem:
\begin{equation*}
\label{eta_range}
\eta_{max}=\max_{\bm{\beta}}\Gamma_e(\bm{\beta}), \text{ subject to: }\;\bm{\beta}^t\Lambda\bm{\beta} \le P_{tot}= P_1 + \ldots + P_M
\end{equation*}
where $\Lambda=diag[h_{s1}^2P_s+\sigma^2,h_{s2}^2P_s+\sigma^2,\cdots,h_{sM}^2P_s+\sigma^2]$. The constraint considered here is total power constraint and corresponds to a larger feasible solution space as compared to individual constraints. Therefore, $\eta_{max}$ calculated here is an over estimate of the maximum value of $\eta$ with individual constraints. The solution of this problem is $\eta_{max}=\mathbf{h_{s,e}}^t\mathbf{\tilde{D}_e}^{-1}\mathbf{h_{s,e}}$, where $\mathbf{\tilde{D}_e}=\frac{1}{P_{tot}}\Lambda+\mathbf{D_e}$.

Along with the solution of the optimal AF secrecy rate in this setting, we also provide upper and lower bounds on the optimal rate. For upper bound, we replace the individual power constraint on each relay by a single total power constraint on $\bm{\beta}$ vector.
 For lower bound, we consider the \textit{zero forcing} solution \cite{109dongHanPetropuluPoor, 113yangLiMaChiang}, where $\bm{\beta}$ values are chosen such that the transmitted signals get canceled at eavesdropper.

\subsubsection{Optimal secrecy rate for the total power constraint}
From equation \eqref{eq:opt1} the total power constraint optimization problem can be expressed in following manner:
\begin{subequations}
	\label{eq:opttotcons}
	\begin{align}
	\max_\eta\;\;\max_{\bm{\beta}}\;\;& \frac{1+SNR_d(\bm{\beta})}{1+\eta} \label{objfn:opttot}\\
	\text{subject to:}\;\; & SNR_e(\bm{\beta})\le \eta,\\ 
	& \bm{\beta}^t\Lambda\bm{\beta} \le P_{tot},\label{subeq:opttotcons3} 
	\end{align}
	
\end{subequations}
One can check that the total power constraint \eqref{subeq:opttotcons3} results in larger feasible solution space as compared to \eqref{eq:beta_cons}. Therefore, the optimal objective function value obtained will upper bound the original individual constraint problem \eqref{eq:optindvcons}. Using the transformation described above, we can reformulate the optimization problem in the following manner:
\begin{subequations}
	\label{eq:totoptrfm}
	\begin{align}
	\max_{\mathbf{v}}\;\;& \mathbf{h_s^tv}\\
	\text{subject to:}\; & \mathbf{v^t}\mathbf{C}_e(\eta)\mathbf{v} \le 1,\label{constr:totoptcons1}\\ 
	& \mathbf{v^tD_Tv}\le 1 \label{constr:totoptcons2}
	\end{align}
\end{subequations}
where $\mathbf{D_T}=diag([\zeta_1,\zeta_2,\cdots,\zeta_M])$,  $\zeta_i=(1+\frac{h_{si}^2P_s+\sigma^2}{h_{id}^2P_{tot}})$. As $\mathbf{D_T}$ is a diagonal matrix with positive diagonal elements, so \eqref{constr:totoptcons2} is a convex constraint. This problem can be solved in a manner similar to the one for the individual power constraint.
}


\subsubsection{Zero forcing for individual constraints ($\eta=0$)}\label{subsubsec:zfc}
In this approach we equate $SNR_e$ to zero while maximizing $SNR_d$. By imposing this extra constraint we are reducing the feasible solution space, therefore, the solution obtained will provide a lower bound to original problem. The optimal secure AF rate problem for this case can be written as:
\begin{subequations}
	\label{zfs}
	\begin{align}
	\max_{\bm{\beta}}&\quad\frac{P_s}{\sigma^2}\frac{\left(\sum_{i=1}^{M} h_{si}\beta_ih_{it}\right)^2}{1 + \sum_{i=1}^{M}(\beta_i h_{it})^2}\label{zfs_obj}\\
	\text{subject to}&\;  \sum\limits_{i=1}^{M} h_{si} \beta_i h_{ie}=0,\\ 
	& -\beta_{i,max} \le\beta_i \le \beta_{i,max}\label{zfs_rel}
	\end{align}
\end{subequations}

We reformulate it as the following quadratic program. 

\begin{pavikp}
	\label{prpn:zf-qp}
	The problem in \eqref{zfs} is equivalent to the following quadratic program which can be solved efficiently
	\begin{equation*}
	\max_{\mathbf{w}}\;\; \mathbf{w^tw},
	\text{ subject to: } {\mathbf{H}}\mathbf{w}=[1 \quad 0]^t, \: \mathbf{H}_\beta\mathbf{w} \le \mathbf{0}
	\end{equation*}
\end{pavikp}
\begin{IEEEproof}
	Please refer to Appendix \ref{appnd2}.
\end{IEEEproof}
\noindent\textit{Remark 1:} As this formulation does not rely on the degradedness assumption, it can be used for general channels also.


\subsection{Scaled Eavesdropper Channels}
\label{subsec:scaledEve}
Scaled eavesdropper scenario is a special case of degraded channel scenario. Here, all the elements of vector $\boldsymbol{\alpha}$ have same value, \textit{i.e.} $\alpha_i = \alpha < 1, i \in \{1, \ldots, M\}$.  Therefore, eavesdropper channel can be expressed as $\bm{h_e}=\alpha \bm{h_t}$. Though the approach discussed in the last section can be used to solve the optimal AF secrecy rate problem corresponding to this network model, here we propose another scheme that better exploits the corresponding channel properties, thus resulting in a more efficient implementation. 
Before going to details, we provide an overview of the proposed solution to the scaled eavesdropper problem.
\begin{enumerate}
	\item Due to the structure of the optimization problem we identify that a common function is present in both numerator and denominator and maximizing that function will result in maximization of the original problem.
	\item Consider a new variable $\omega_i=h_{id}\beta_{i}$, its upper bound $\omega_{i,max}=h_{id}\beta_{i,max}$, and a new parameter $g_{si}=\sqrt{\frac{P_s}{\sigma^2}}h_{si}, i \in \{1,2,\ldots,M\}$. Now we relaxed the derived optimization problem by replacing all the individual constraint by a single total constraint which can be solved efficiently.
	\begin{align*}
		\textrm{ \textbf{Original}:}& \max f(\boldsymbol{\omega})=\left(\sum\limits_{i=1}^M g_{s,i}\omega_i\right)^2 \text{subject to: } \omega_i \le \omega_{i,max}\,i \in\{1,2,\ldots,M\} \\
		\textrm{ \textbf{Relaxed:}}&	\max \left(\sum\limits_{i=1}^M g_{s,i}\omega_i\right)^2 \text{subject to: }\sum\limits_{i=1}^{M}\omega_i^2=r^2
	\end{align*}
	Here, $r^2 \in [0,||\boldsymbol{\omega}||^2]$
	\item  If the solution of this relaxed problem satisfy individual constraints, then we have solved the problem optimally.
	\item  If the solution of that total constraint problem violates individual constraints, then we have to impose the individual constraint criteria to obtain the solution. In other words, some of the variables have reached their boundary value and others are still inside constraint region. The question is how to identify the variables that have reached their respective upper bounds?
	
	Let us begin with the obvious scenario, where $r^2=||\boldsymbol{\omega}_{max}||^2$. The solution is trivial \textit{i.e.}  $\omega_i=\omega_{i,max}, \; \forall \{1,2,\ldots,M\}$.  Now, if we want to reduce $\delta$ from all the variables, how should we proceed so that objective is least effected. Consider, $0\le\theta_i\le1$ and $\sum\limits_{i}^{}\theta_i=1$. If we reduce $\delta\theta_i$ from each variable, then for  $\delta \to 0$, neglecting $\delta^2$, the rate of change in objective function with respect to $r^2$ can be written as:
	\[\der[f]{r^2} =\sum\limits_{i}^{}\pder[f]{\omega_i}/\pder[r^2]{\omega_i}=\sum\limits_{i}^{}\theta_i\frac{g_{s,i}}{\omega_{i,max}}\ge \frac{g_{s,i^*}}{\omega_{i^*,max}} \text{where } i^* =\arg \min_i \frac{g_{s,i}}{\omega_{i,max}} \]
	Therefore, we should reduce the variable $\omega_{i^*}$, to keep the rate of decrement of the objective function minimum. 
We can keep reducing $\omega_{i^*}$ only till $\frac{g_{s,i^*}}{\omega_{i^*}}=\frac{g_{s,j^*}}{\omega_{j^*,max}}$, where $j^* =\arg \min_{j \ne i} \frac{g_{s,j}}{\omega_{j,max}}$. From that point onwards variables $\omega_{i^*}$ and $\omega_{j^*}$ both has least effect on objective function among all variables and, therefore,  we keep reducing them jointly till we reach $\frac{g_{s,k^*}}{\omega_{k^*,max}}$, where $k^* =\arg \min_{k \ne i,j} \frac{g_{s,k}}{\omega_{k,max}}$ and it continues in similar manner. 
	
	 From the above discussion one can conclude that an ordering among the variables are possible depending on their contribution towards objective function. Variables that contributes more towards objective function have higher tendency of violating individual constraints as compared to other variables. So, we sort the variables in descending order based on there $\frac{g_{s,i}}{\omega_{i,max}}$ value and re-index them in following manner:
	 \[\frac{g_{s,(1)}}{\omega_{(1),max}} \ge \frac{g_{s,(2)}}{\omega_{(2),max}} \ge \cdots \ge \frac{g_{s,(M)}}{\omega_{(M),max}}\]
	 Due to re-indexing, we can say that $\omega_{(i)}$ has higher tendency of violating individual constraint than $\omega_{(i+1)}$. Now, we can divide the range $[0,||\boldsymbol{\omega}_{max}||^2]$ based on this ordering. An interval $[r_m^2,r_{m+1}^2]$ is defined in following manner:
	 \[r_m^2=\sum\limits_{i=1}^{m}\omega_{(i),max}^2+\frac{\omega_{(m),max}^2}{g_{s,(m)}^2}\sum\limits_{i=m+1}^{M}g_{s,(i)}^2
	 \]
	In simple words, within this interval variables indexed by $(1)$ to $(m)$ has reached their upper bounds and we need to solve the total constraint problem for rest of the variables.
	\item We argue that the solution for each of these subproblems within an interval can be obtained by solving a one-dimensional maximization problem.
	\item We choose the solution that maximizes the original objective function over those sub-intervals. 
\end{enumerate}

We rewrite the achievable AF secrecy rate in terms of $\bm{\omega}$:
\begin{equation*}
R_s(\bm{\omega})=\frac{1}{2}\log\left[\frac{1+\varrho_1\Psi(\bm{\omega})}{1+\varrho_2\Psi(\bm{\omega})}\right],
\end{equation*}
where $\Psi(\mathbf{\omega})=\left(\sum_{i=1}^M g_{si}\omega_i\right)^{\!\!2}$, $\varrho_1=\frac{1}{1+r^2}$, and $\varrho_2=\frac{\alpha^2}{1+\alpha^2r^2}$ with $\sum_{i=1}^{M}\omega_i^2=r^2$. So, in this setting the problem in \eqref{eq:opt1} is:
\begin{equation}
\label{eqn:sclEveProb}
R_s(P_s) =  \max_{\bm{\omega}} R_s(\bm{\omega})
\end{equation}

In the rest of this subsection we provide an approach to efficiently compute the optimal solution of the problem in \eqref{eqn:sclEveProb} and then summarize it in an algorithm.

Note that a non-zero secrecy rate can be obtained only if $\alpha <1$. Thus, $\varrho_1 > \varrho_2$. This implies that $\frac{1+\varrho_1\Psi(\bm{\omega})}{1+\varrho_2\Psi(\bm{\omega})}$ is maximized when $\Psi(\bm{\omega})$ is correspondingly maximized. To maximize $\Psi(\bm{\omega})$, we formulate the following problem:
\begin{equation}
\label{opt:scl_eve}
\max \left(\sum\limits_{i=1}^M g_{si}\omega_i\right)^2 \text{subject to} \sum\limits_{i=1}^{M}\omega_i^2=r^2
\end{equation}

It is easy to prove that the optimal $\bm{\omega}(r)$ vector for this problem is equal to $\bm{\omega}(r) = \frac{\mathbf{g_s}}{||\mathbf{g_s}||}r$. By replacing this in the secrecy rate equation \eqref{eqn:sclEveProb} along with the sum constraint we formulate the following optimization problem in $r$:
\begin{align}
\label{opt:onedim}
\max_r \quad\frac{1+\frac{||\mathbf{g_s}||^2r^2}{1+r^2}}{1+\frac{\alpha^2||\mathbf{g_s}||^2r^2}{1+\alpha^2r^2}}
\end{align} 

\begin{pavikl}
	\label{lem:optonedim}
	The optimal solution to the problem in \eqref{opt:onedim} is $r^*=(\sqrt{\alpha\sqrt{1+||\mathbf{g_s}||^2}})^{-1}$.
\end{pavikl}
\begin{IEEEproof}
	By differentiating the objective function of \eqref{opt:onedim} with respect to $r$ and then it equating to $0$, we get the solution.
\end{IEEEproof}

Now, if  $\frac{\mathbf{g_s}}{||\mathbf{g_s}||}r^*$ satisfy individual constraints of $\omega_i$ then this is the optimal solution, else we must consider the next step that considers scenarios where the solution obtained using above method violates any of the $\omega_i$'s constraints.

Arrange $\omega_i$'s according to $\frac{g_{si}}{\omega_{i,max}}$ and denote them as $\omega_{(1)} \ge \ldots \ge \omega_{(M)}$. The $g_{si}$ are ordered accordingly with the ordered values denoted as $g_{s,(i)}$. Then, we divide the feasible set of the problem in \eqref{opt:scl_eve} in $M$ non-overlapping subsets, $[r_m, r_{m+1}]$, $m \in \{0, \ldots, M-1\}$. The interval that results in the largest value of the problem in \eqref{eqn:sclEveProb} solved over it as discussed below, is chosen as the one that results in the optimal solution of \eqref{eqn:sclEveProb} over all intervals and the corresponding optimal vector of scaling factors $\bm{\beta}^*$. The correctness of this approach follows from an argument similar to the one in \cite[Lemma 1]{109jingJafarkhani}.

For the ease of exposition, define $p_m = \sum_{i=1}^{m}g_{s,(i)}\omega_{(i),max}$, $q_m = \sum_{i=1}^{m}\omega_{(i),max}^2$, and $s_m = \sum_{i=m+1}^{M}g_{s,(i)}^2$.


The intervals can be calculated as follows. The first interval is $[0,r_1]$, where $r_1$ corresponds to the value at which the first variable (after ordering) is equal to its upper bound. Therefore, $\bm{\omega}^*= \frac{\mathbf{g_s}}{||\mathbf{g_s}||}r_1=\frac{\mathbf{h_s}}{||\mathbf{h_s}||}r_1$ and $\omega_{(1),max}=\frac{g_{s,(1)}}{||\mathbf{g_s}||}r_1$. Thus, $r_1 = \frac{||\mathbf{g_s}||}{g_{s,(1)}}\omega_{(1),max}=\sqrt{(s_1/g_{s,(1)}^2)\omega_{(1),max}^2\!\!+\!q_1}$. Similarly, $r_m=\sqrt{\!(s_m/g_{s,(m)}^2\!)\omega_{(m),max}^2\!\!\!+\!q_m}$, $m \!\in\! \{1,\ldots,\!M\!-1\}$ and $r_M=q_M$.

\noindent\textit{Remark 2:} If $0 \le r \le r_1$, then we can obtain the optimal scaling vector using the solutions of \eqref{opt:scl_eve} and \eqref{opt:onedim}.

Consider evaluating \eqref{eqn:sclEveProb} over the interval $[r_m,r_{m+1}]$. In this stage the first $m$ ordered variables have reached their respective boundaries. These variables are replaced by their respective upper bounds. For the rest of the variables we search for their respective optimum values as discussed next.

The part of the objective function in \eqref{opt:scl_eve} depending on variables $[\omega_{(m+1)},\ldots,\omega_{(M)}]$ can be upper bounded as:
\begin{align*}
\mathbf{g}_{s,\{(m+1),\cdot,(M)\}}^t\bm{\omega}_{\{(m+1),\cdot,(M)\}} & \le \lambda_m ||\mathbf{g}_{s,\{(m+1),\cdot,(M)\}}||\\
\text{for all }||\bm{\omega}_{\{(m+1),\cdot,(M)\}}||&=\sqrt{r^2-q_m},
\end{align*}
where $\mathbf{u}_{\{(m+1),\cdot,M\}}:=[u_{(m+1)},\cdots,u_{(M)}]^t$ and $\lambda_m > 0$. In other words, the objective function is maximized when the vector variables $[\omega_{(m+1)},\ldots,\omega_{(M)}]^t$ lies in the direction of $[g_{s,(m+1)},\cdots,g_{s,(M)}]^t$ vector. Therefore, we have to find the optimal scaling factor $\lambda_m$ for those variables. Replacing the first $m$ ordered variables by their respective upper bounds and rest of them by a scaled vector of $\mathbf{g}_{s,\{(m+1),\cdot,(M)\}}$, we obtain the following optimization problem in terms of $\lambda_m$.
\begin{equation*}
\max_{\lambda_m: \lambda_m > 0} \frac{1+\gamma_s\frac{(p_m+s_m\lambda_m)^2}{1+q_m+s_m\lambda_m^2}}{1+\gamma_s\frac{\alpha^2(p_m+s_m\lambda_m)^2}{1+\alpha^2q_m+\alpha^2s_m\lambda_m^2}}    
\end{equation*}

The solution to the above problem can be calculated by solving the following degree-4 polynomial equation:          
\begin{multline}
\label{eq:polynm}
\frac{(1+q_m)(1+\alpha^2q_m)}{s_m^2\alpha^2(1+\gamma_ss_m)}
-\frac{p_m(\gamma_sp_m^2\alpha^2+2q_m\alpha^2+\alpha^2+1)}{s_m^2\alpha^2(1+\gamma_ss_m)}\lambda_m\\
-\frac{3p_m^2\gamma_s}{s_m(1+\gamma_ss_m)}\lambda_m^2
-\frac{p_m(2+3\gamma_ss_m)}{s_m(1+\gamma_ss_m)}\lambda_m^3
-\lambda_m^4=0
\end{multline}

The coefficients of this equation are all negative except the first one. Using the \textit{Descartes' rule of sign} there is only one positive root as there is only one variation in sign of the coefficients. Given that $0 < \lambda_m$, it is the desired solution.
Note that, if $\lambda_m > w_{(m+1),max}/g_{s,(m+1)}$, then the solution will not belong to predefined sub-interval $(\because r_{m+1}^2={\!(s_{m+1}/g_{s,(m+1)}^2\!)\omega_{(m+1),max}^2\!\!\!+\!q_{m+1}}<{\!\lambda_ms_{m+1}\!\!\!+\!q_{m+1}})$.\\
We summarize the whole procedure in Algorithm \ref{algo1}.
\vspace{-0.1in}
\begin{algorithm}[!h]
	\caption{Algorithm for Scaled Eavesdropper Channel} 
	\textbf{Input:} $\alpha$, $\bm{g_s}$, $\bm{\omega}_{i,max}$, $P$, $\sigma^2$\\
	\textbf{Output:} $\bm{\omega}^*$
	\begin{algorithmic}[1]
		\STATE Calculate $r^*=\frac{1}{\sqrt{\alpha\sqrt{1+||\mathbf{g_s}||^2}}}$
		and	evaluate $\bm{\omega}=\frac{\mathbf{g_s}}{||\mathbf{g_s}||}r^*$.\\
		\IF{$\omega_i \le \omega_{i,max}$, $i \in \{1, \ldots, M\}$}
		\STATE	$\bm{\omega}^* = \bm{\omega}$ (Lemma \ref{lem:optonedim}) and exit.
		\ELSE
		\STATE 	Sort $\omega_i$s according to $\frac{g_{si}}{\omega_{i,max}}$ in the descending order.
		\FOR{$m=1$ to $M-1$}
		\STATE	Solve polynomial equation \eqref{eq:polynm} to calculate the $\lambda_m$. 
		\STATE Evaluate $\bm{\omega}_m$ using $\lambda_m$ corresponding to interval $[r_m,r_{m+1}]$ starting from first $i$ for which $\lambda_i < w_{(i+1),max}/g_{s,(i+1)}$. Denote that $i$ as $i_0$.		
		\ENDFOR	
		\STATE	$\bm{\omega}^* = \bm{\omega}_m^*$, $\bm{\omega}_m^*$ maximizes the solution of \eqref{eqn:sclEveProb} over all $\bm{\omega}_m$, $m \in \{i_0,\ldots,M-1\}$.
		\ENDIF			
	\end{algorithmic}
	\label{algo1} 
\end{algorithm}           

\subsection{Symmetric Network}
\label{subsec:symNet}
A special case of the general $s-t$ and $s-e$ diamond networks introduced above is the \textit{symmetric} setting, where $h_{si} = h_s, h_{it} = h_t$, $h_{ie} = h_e$ and {$P_i=P_R, i \in \{1, \ldots, M\}$.} In this setting, for a given vector of scaling factors, $\bm{\beta} = (\beta_1, \ldots, \beta_M)$, the achievable AF secrecy rate is:
\begin{equation*}
R_s(\bm{\beta}) \! = \! \frac{1}{2}\log\!\left[\left(\!\!1\!+\!\frac{\gamma_s^{'}\left( \sum\limits_{i=1}^{M}\beta_i\!\right)^{\!\!2}}{\nu+\sum\limits_{i=1}^{M}\beta_i^2}\right)\!\bigg/\!\! \left(\!\! 1\!+\!\frac{\gamma_s^{'}\left(\sum\limits_{i=1}^{M}\beta_i\!\right)^{\!\!2}}{\mu+\sum\limits_{i=1}^{M}\beta_i^2}\right)\right]\!\!,
\end{equation*}
where
$\gamma_s^{'}=\gamma_s h_s, \nu=h_t^{-2}$, and $\mu=h_e^{-2}$. Note that $R_s(\bm{\beta})$ is non-zero only if $\nu<\mu$, \textit{i.e.} $h_t>h_e$. Thus, using \eqref{eq:opt1}, the optimal secure AF rate problem in this setting is:
\begin{equation}
\label{eqn:symNetProb}
R_s(P_s) =  \max_{\bm{\beta}} R_s(\bm{\beta})
\end{equation}

\begin{pavikp}
\label{prpn:symNet}
For the symmetric $M$-relay diamond network, the optimum scaling factors for all relays are equal, \textit{i.e.} $\beta^*_1=\beta^*_2=\dots=\beta^*_M = \beta^*$.
\end{pavikp}
\begin{IEEEproof}
Please refer to Appendix \ref{appnd1}.
\end{IEEEproof}

\begin{pavikc}
\label{cor:symNet}
For symmetric $M$ relay diamond network, the optimal value of the scaling factor for each relay is:
\begin{equation*}
\beta^* = \begin{dcases}
            \bigg[\frac{\sigma^2}{P_R M^2h_t^2 h_e^2}\bigg]^{1/4} \beta_{max}^{1/2}, \quad \frac{\sigma^2}{P_R \beta_{max}^2} < M^2h_t^2 h_e^2 \\
            \beta_{max}, \qquad \mbox{o.w.}
            \end{dcases}
\end{equation*}
\end{pavikc}
\begin{IEEEproof}
Using Proposition~\ref{prpn:symNet}, the optimal AF secrecy rate problem in \eqref{eq:opt1} is reduced in this case to the following single parameter optimization problem:
\begin{equation*}
\max_{\beta: 0 \le \beta^2 \le \beta_{max}^2}\left( 1+\frac{P_s}{\sigma^2} \frac{M^2\beta^2 h_s^2 h_t^2}{1 + M\beta^2 h_t^2}\right)\bigg/\left( 1+\frac{P_s}{\sigma^2} \frac{M^2\beta^2 h_s^2 h_e^2}{1 + M\beta^2 h_e^2}\right)
\end{equation*} 
With the parameters: $h_{sr}=\sqrt{M}h_s$, $h_{rd}=\sqrt{M}h_t$, and $h_{re}=\sqrt{M}h_e$, this problem reduces to the single relay AF secrecy rate maximization problem, addressed in Lemma~\ref{lem:single-rel} below.
\end{IEEEproof}

\begin{pavikl}
\label{lem:single-rel}
The optimal AF scaling factor for single AF relay network is:
\begin{equation*}
\beta_{AF} = \begin{dcases}
				\bigg[\frac{\sigma^2}{P_R h_{rd}^2 h_{re}^2}\bigg]^{1/4} \beta_{max}^{1/2}, \quad \frac{\sigma^2}{P_R \beta_{max}^2} < h_{rd}^2 h_{re}^2 \\
				\beta_{max}, \quad \mbox{o.w.}
			\end{dcases}
\end{equation*}
\end{pavikl}
\begin{IEEEproof}
Please refer to Appendix \ref{appnd4}.
\end{IEEEproof}

\textit{Remark 3:} One can obtain an upper bound on secrecy rate for degraded channel scenario (\ref{subsec:degradedEve}) by using the result of symmetric channel scenario. For example, if we form a $M$ relay symmetric network by considering $h_{s}=\max\{\bf{h_s}\}$, $h_{t}=\max\{\bf{h_t}\}$ and $h_e=\min\{\bf{h_e}\}$, where $\max\{\bf{x}\}$ and $\min\{\bf{x}\}$ denotes the maximum and minimum element of vector $\bf{x}$, respectively.
\section{Results}
\label{sec:res}
To evaluate the performance of the scheme proposed in subsection~\ref{subsec:degradedEve}, consider a network whose main channel gains are sampled from a Rayleigh distribution with parameter 0.5. The degraded channels for eavesdroppers is obtained by multiplying a factor ($\sim Uniform[0,1]$) with the relay to destination channel gain. We average the results of 1000 such network instances while plotting the graphs.

\begin{figure}[!t]
	\centering
	\includegraphics[width=3.4in]{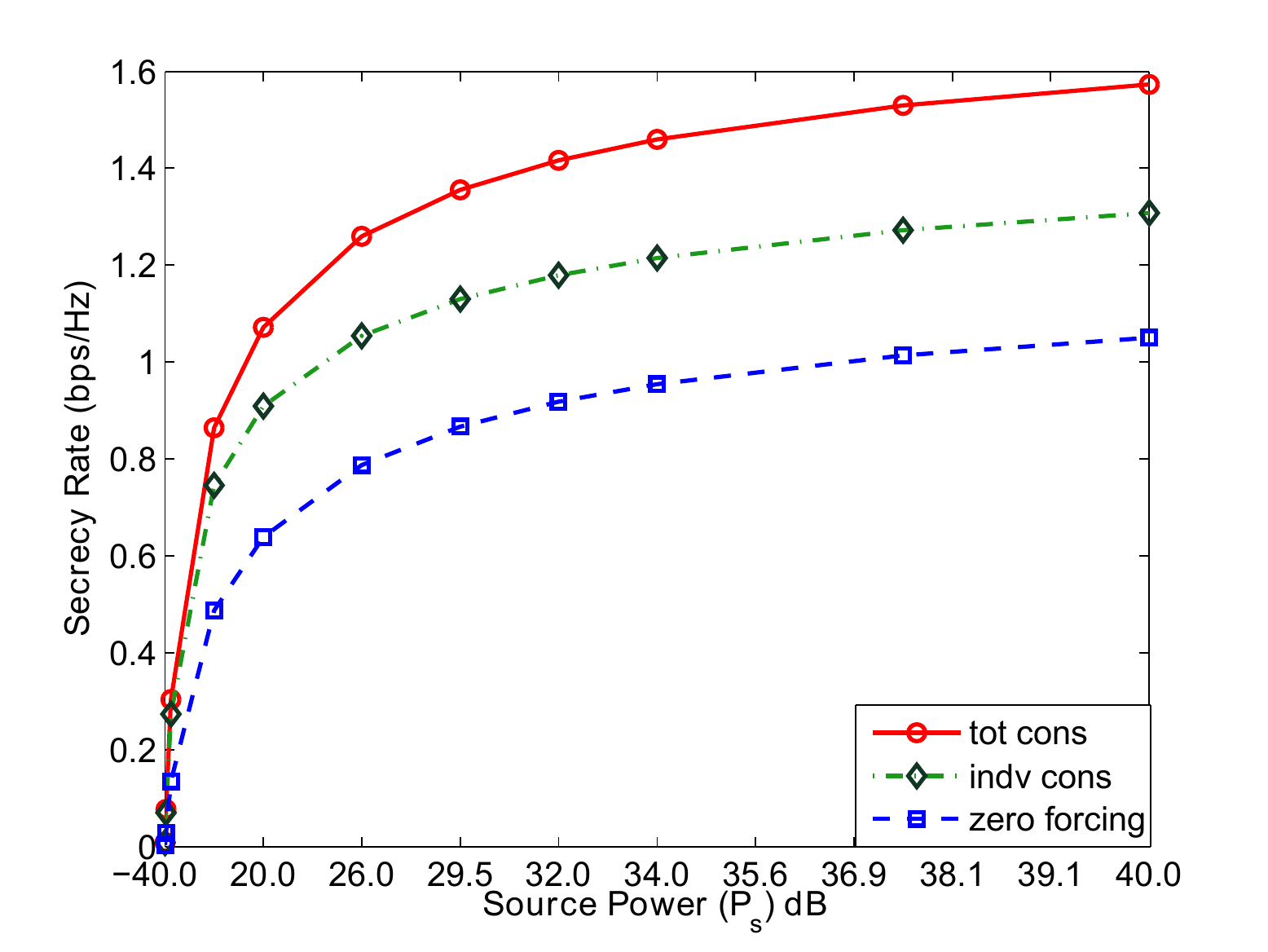}
	\vspace{-0.15in}
	\caption{Plot of the optimal secrecy rate ($R_s^*$) versus source power ($P_s$) for five relay node diamond network using the scheme proposed in subsection~\ref{subsec:degradedEve}. Here we consider $P_r=5$ and $\sigma^2=1$.}
	\vspace{-0.2in}
	\label{fig_secrate}
\end{figure}


In Figure \ref{fig_secrate} we plot the variation of the optimal AF secrecy rate with respect to the source power ($P_s$) for the degraded eavesdropper channels. The zero-forcing lower bound and the total power constraint upper bounds are also plotted.

 
\section{Conclusion}
\label{sec:concl}
The problem of maximizing the rate of secure unicast transmission over a network of parallel AF relays in the presence of an eavesdropper is considered. Unlike previous work that provides iterative algorithms with suboptimal performance without any performance and convergence guarantees, our work provides either a closed-form expression or polynomial-time algorithm to compute the optimal secure AF rate for three specific network models considered. In future, we plan to investigate optimum secrecy rate of general AF networks.

\appendices
\section{Proof of Proposition \ref{prpn:unimod}}\label{appnd3}
\begin{IEEEproof}
	In optimization problem \eqref{opt_indv}, the individual constraints \eqref{eq:consindv} are independent of $\eta$. If a vector $\mathbf{x_1}$ satisfies the constraint \eqref{eq:evecons} for $\eta_1$, then $\mathbf{x_1}$ will satisfy it for $\eta_2>\eta_1$. 
	$1\ge \mathbf{x_1^tC}_e(\eta_1)\mathbf{x_1}=\frac{(\hh^\mathbf{t}\mathbf{x_1})^2}{\eta_1}\frac{P_s}{\sigma^2}+\mathbf{x_1^t}\mathbf{\overline{D}_e}\mathbf{x_1}>\frac{(\hh^\mathbf{t}\mathbf{x_1})^2}{\eta_2}\frac{P_s}{\sigma^2}+\mathbf{x_1^t}\mathbf{\overline{D}_e}\mathbf{x_1}=\mathbf{x_1^tC}_e(\eta_2)\mathbf{x_1}$, here $\mathbf{\overline{D}_e}=\mathbf{I}-\Dh_e$. In fact, as $\eta$ increases feasible set for \eqref{eq:evecons} also increases. As a result the feasible set for optimization problem \eqref{opt_indv} also increases. Therefore, $(\mathbf{h_s^tv}^*(\eta_1))^2\le (\mathbf{h_s^tv}^*(\eta_2))^2$. Recall that, $\eta=0$ corresponds to zero-forcing solution (Sec. \ref{subsubsec:zfc}) and, therefore, $f(\eta)$ increase with $\eta$ initially. 
	But later constraint \eqref{constr:totoptcons2} become active and constrain the feasible set. As a result, the numerator of $f(\eta)$ remains fixed, whereas, denominator increases with $\eta$. This results in decrease of $f(\eta)$.
\end{IEEEproof}

\section{Proof of Proposition \ref{prpn:zf-qp}}
\label{appnd2}
\begin{IEEEproof}
	It can be shown that the optimal solution does not change if we rewrite the objective function as $\min\;\frac{1 +\sum_{i=1}^{M}(\beta_i h_{id})^2}{\left( \sum_{i=1}^{M} h_{si}\beta_ih_{id}\right)^2}$, because $ \sum_{i=1}^{M} h_{si}\beta_ih_{id}\ne 0$. If we consider a new variable $v$ such that $\sum_{i=1}^{M} h_{si}\beta_ih_{id}=v$, then in terms of variable vector $\mathbf{w}=[\frac{\bm{\omega}}{v}^{t},\frac{1}{v}]^t$, we can easily write a quadratic optimization problem.
	The denominator of the objective function \eqref{zfs_obj} can be written as a constraint in terms of this variable vector as, $[\mathbf{h}_s^\mathbf{t},0]\mathbf{w}=1$. The eavesdropper constraint can be expressed as $[\hh^t,0]\mathbf{w}=0$. We combine them as matrix $\mathbf{H}$ and write as equality constraint.
	$\mathbf{H}_\beta$ can be obtained by rearranging the following constraint: $-h_{d,i}\beta_{i,max}w_{M+1} \le w_i \le h_{d,i}\beta_{i,max}w_{M+1},\forall i$.
\end{IEEEproof}


\section{Proof of Proposition \ref{prpn:symNet}}
\label{appnd1}
\begin{IEEEproof}
Taking derivative of $R_s$ with respect to $\beta_i$, we get
\begin{align*}
2\gamma'_s & \Bigg\{ \Big( \sum\limits_{i=1}^{M}\beta_i^2-\beta_i\Big( \sum\limits_{j=1}^{M}\beta_j\Big) \Big)\Big( 2 \sum\limits_{i=1}^{M}\beta_i^2+\nu+\mu\Big) \\
-&\Big( \sum\limits_{i=1}^{M}\beta_i^2\Big) ^2-\gamma'_s\beta_i\Big( \sum\limits_{j=1}^{M}\beta_i\Big)^3+\nu\mu\Bigg\}\Big( \sum\limits_{j=1}^{M}\beta_i\Big) \Big( \mu-\nu\Big) =0
\end{align*}

Now, as $\left( \sum\limits_{j=1}^{M}\beta_i\right)=0$ is not the optimal solution and $\mu\ne\nu$, hence
\begin{multline}
\label{gammaeq1}
\left( \sum\limits_{i=1}^{M}\beta_i^2-\beta_i\left( \sum\limits_{j=1}^{M}\beta_j\right) \right)\left( 2 \sum\limits_{i=1}^{M}\beta_i^2+\nu+\mu\right) -\left( \sum\limits_{i=1}^{M}\beta_i^2\right) ^2\\
-\gamma'_s\beta_i\left( \sum\limits_{j=1}^{M}\beta_i\right)^3+\nu\mu =0
\end{multline}
Similar equation can be obtained if we take derivative with respect to $\beta_j,\;j\ne i \text{ and } j \in \{1,2,\dots,M\}$. Subtracting this equation from equation \eqref{gammaeq1} we get
{\small \begin{equation}
		(\beta_i-\beta_j)\left( \sum\limits_{j=1}^{M}\beta_j\right)\left\lbrace\left( 2 \sum\limits_{i=1}^{M}\beta_i^2+\nu+\mu\right)+\gamma'_s\left( \sum\limits_{j=1}^{M}\beta_i\right)^3\right\rbrace=0 
		\end{equation}}
Hence, $\beta_i=\beta_j,\;i,j\in\{1,2,\dots,M\}$
\end{IEEEproof}  

\section{Proof of Lemma \ref{lem:single-rel}}
\label{appnd4}
For single relay network the problem is:
\begin{equation*}
\beta_{AF} = \argmax_{\beta}\left(1+\frac{P_s}{\sigma^2} \frac{h_{sr}^2 \beta^2 h_{rd}^2}{1+ \beta^2 h_{rd}^2}\right)\bigg/ \left( 1 +\frac{P_s}{\sigma^2} \frac{h_{sr}^2 \beta^2 h_{re}^2}{1 + \beta^2 h_{re}^2}\right) 		
\end{equation*}
Differentiating the objective function with respect to $\beta$ and setting the derivative equal to zero, we get for the optimum $\beta$ value
\begin{equation*}
\beta^4 = \bigg[\frac{\sigma^2}{P_R h_{rd}^2 h_{re}^2}\bigg] \beta_{max}^2
\end{equation*}
Further, given that the maximum value that $\beta$ can take is bounded from above by $\beta_{max}$ and the monotonic increasing nature of the objective function from $\beta = 0$ to $\beta = \beta_{max}$, we obtain the desired result.


\begin{thebibliography}{99}
\bibitem{wyner} A.~Wyner, ``The Wire-tap channel,'' \textit{Bell Sys. Tech. Journal}, vol.~54, January 1975.

\bibitem{078cheongHellman} S.~L.-Y.-Cheong and M.~Hellman, ``The Gaussian wire-tap channel,'' \textit{IEEE Trans. Inform. Theory}, vol. IT-24, July 1978.

\bibitem{105paradaBlahut} P.~Parada and R.~Blahut, ``Secrecy capacity of SIMO and slow fading channels,'' \textit{Proc. IEEE ISIT}, Adelaide, Australia, September 2005.

\bibitem{110khistiWornell} A.~Khisti and G.~W. Wornell, ``Secure transmission with multiple antennas i:  The MISOME wiretap channel,'' \textit{IEEE Trans. Inform. Theory}, vol. IT-56, July 2010.

\bibitem{108liuMaric} R.~Liu, I.~Maric, P.~Spasojevic, and R.~D.~Yates, ``Discrete memoryless interference and broadcast channels with confidential messages: Secrecy capacity regions,'' \textit{IEEE Trans. Inform. Theory}, vol. IT-54, June 2008.

\bibitem{108khisti_thesis} A. Khisti, \textit{Algorithms and Architectures for Multiuser, Multiterminal,
and Multilayer Information-theoretic Security.} Ph.D. thesis, Massachusetts Institute of Technology, 2008.

\bibitem{108liangPoorShamai} Y.~Liang, H.~V. Poor, and S.~Shamai, ``Secure communication over fading channels,'' \textit{IEEE Trans. Inform. Theory}, vol. IT-54, June 2008.

\bibitem{108gopalaLaiGamal} P.~Gopala, L.~Lai, and H.~El~Gamal, ``On the secrecy capacity of fading channels,'' \textit{IEEE Trans. Inform. Theory}, vol. IT-54, October 2008.

\bibitem{108laiGamal} L.~Lai and H.~El~Gamal, ``The relay-eavesdropper channel: Cooperation for secrecy,'' \textit{IEEE Trans. Inform. Theory},  vol. IT-54, September 2008.

\bibitem{109dongHanPetropuluPoor} L.~Dong, Z.~Han, A.~Petropulu, and H.~Poor, ``Amplify-and-Forward based cooperation for secure wireless communications,'' \textit{Proc. IEEE ICASSP}, Taipei, Taiwan, April 2009.

\bibitem{110zhangGursoy} J.~Zhang and M.~C. Gursoy, ``Relay beamforming strategies for physical-layer security,'' \textit{Proc. IEEE CISS}, Princeton, NJ, March 2010.

\bibitem{110zhangGursoy2} J.~Zhang and M.~C. Gursoy, ``Collaborative relay beamforming for
secrecy,'' \textit{Proc. IEEE ICC}, Cape Town, South Africa, May 2010.

\bibitem{113yangLiMaChiang} Y.~Yang, Q.~Li, W.-K. Ma, J.~Ge, and P.~C.~Chiang, ``Cooperative secure beamforming for AF relay networks with multiple eavesdroppers,'' \textit{IEEE Sig. Proc. Lett.}, vol.~20, January 2013.

\bibitem{109jingJafarkhani} Y.~Jing and H.~Jafarkhani, ``Network beamforming using relays with perfect channel information,'' \textit{IEEE Trans. Inform. Theory}, vol. IT-55, June 2009.

\bibitem{111agnihotriJaggiChen} S.~Agnihotri, S.~Jaggi, and M.~Chen, ``Amplify-and-Forward in wireless relay networks,'' \textit{Proc. IEEE ITW}, Paraty, Brazil, October 2011.

\bibitem{115sarmaAgnihotriKuri} S.~Sarma, S.~Agnihotri, and J.~Kuri, ``Secure transmission in amplify-and-forward diamond networks with a single eavesdropper," arXiv:1504.03149.

\bibitem{ursdiggavi} U.~Niesen and Suhas N. Diggavi, ``The approximate capacity of the Gaussian N-relay diamond network." \textit{IEEE Transactions on Information Theory}, 59.2 (2013): 845-859.


\bibitem{Chong2013} E.~K. Chong and S.~H. Zak, \textit{An Introduction to Optimization}. John Wiley \& Sons, 2013.

%
%







\end{thebibliography}
\end{document}